\documentclass[12pt]{article}

\usepackage{latexsym}

\font\tenmsbm=msbm10 scaled 1200
\font\sevenmsbm=msbm9
\newfam\msbmfam
\textfont\msbmfam=\tenmsbm
\scriptfont\msbmfam=\sevenmsbm

\makeatletter
\@addtoreset{equation}{section}
\makeatother

\newcommand{\eref}[1]{(\ref{#1})}

\def\be{\begin{equation}}
\def\ee{\end{equation}}
\def\ba{\begin{eqnarray}}
\def\ea{\end{eqnarray}}
\def\bet{\begin{tabular}}
\def\eet{\end{tabular}}

\def\ve{\varepsilon}

\def\yh{\hat y}

\catcode`@=11

\begin{document}

\begin{titlepage}

\begin{flushright}
Preprint DFPD 04/TH/02\\
February 2004\\
\end{flushright}

\vspace{0.5truecm}

\begin{center}

{\Large \bf Intersecting $M2$-- and $M5$--branes}

\vspace{2.5cm}

K. Lechner
\vspace{2cm}

 {
\it Dipartimento di Fisica, Universit\`a degli Studi di Padova,

\smallskip

and

\smallskip

Istituto Nazionale di Fisica Nucleare, Sezione di Padova,

Via F. Marzolo, 8, 35131 Padova, Italia
}

\vspace{2.5cm}

\begin{abstract}

If an $M2$--brane intersects an $M5$--brane the canonical Wess--Zumino
action is plagued by a Dirac--anomaly, i.e. a non--integer change
of the action under a change of Dirac--brane. We show that this
anomaly can be eliminated at the expense of a gravitational anomaly 
supported on the intersection manifold. Eventually we check that the 
last one is cancelled by the anomaly produced by the fermions present. 
This provides a quantum consistency check of these intersecting 
configurations.

\vspace{0.5cm}

\end{abstract}

\end{center}

\vfill
\hrule width 5.2cm
\vskip2.mm
{\small \noindent e-mail: kurt.lechner@pd.infn.it}
\end{titlepage}

\newpage

\baselineskip 6 mm


\section{Introduction and Summary}

An $M2$-- and an $M5$--brane form an electromagnetically dual
pair of branes in eleven dimensions, and
since $3+6-11$ is a negative number their worldvolumes have 
generically an empty intersection. 
However, for exceptional brane configurations it can happen that 
their intersection is non empty. 
Their intersection manifold $\Sigma\equiv M2 \cap M5$ can then be a manifold 
of dimensions $d=0,1,2$ or $3$. An analysis of the 
{\it quantum}--consistency of such intersections, 
a special case of so called {\it non--transversal} intersections between
two generic manifolds \cite{BT,CY}, is the main topic of  
this letter. The are two types of quantum inconsistencies we will have
to worry about: 1) gravitational ABBJ--anomalies and 2) Dirac--anomalies i.e.
(non--integer) changes of the action under a change of the Dirac--brane. As
we will see these two types of anomalies are intimately related.

The relevance for $M$--theory of the exceptional configurations considered
in this letter, stems from the fact that eleven--dimensional supergravity
admits indeed classical susy--preserving solutions, that can be interpreted 
as an $M2$--brane intersecting with an $M5$--brane \cite{Izq}--\cite{Ohta}.
These solutions are typically localized only in the {\it common} transverse
directions of the two branes, i.e. the currents of the branes are 
$\delta$--functions only in the common transverse coordinates. There exist
also susy--preserving (implicit) solutions where one of the two branes is
fully localized and the other is localized only in the common transverse 
directions \cite{Tsey2,Local}. In absence of a complete classification
of all possible solutions, the existence of solutions where both branes are 
fully localized is still an open question. 

Nevertheless, in this letter we assume that from a quantum point of view 
both branes are fully localized. The fundamental reason for this assumption
is that since $M2$ and $M5$ are dual
objects, only if both currents are $\delta$--functions on the corresponding 
worldvolumes there exists a consistent minimal coupling of the 
branes to each other, because only in this case the charge is {\it locally} 
integer and, therefore, the Dirac--brane unobservable, \cite{ZW,LM}. The 
secondary reason is that 
only branes with a $\delta$--like support represent a universal type of
charge distribution. 
The situation is similar to the $D=4$ Julia--Zee dyons \cite{JZ}. These
dyons represent semi--classical solutions of the Georgi--Glashow model, whose
magnetic charge is fully localized (point--like localization) while their 
electric charge is smeared out. Nevertheless, at the level of second quantized
quantum field theory the Julia--Zee dyons appear with fully localized 
magnetic and electric charges \cite{FM}. 

While this point deserves clearly further investigation, here we take a 
pragmatic point of view and assume that both branes, and therefore also their 
possible intersection, carry well--defined worldvolumes and that the 
associated currents are $\delta$--functions supported on those worldvolumes.

For generic configurations ($\Sigma=\emptyset$) the two branes are at
a finite non--vanishing distance and their classical dynamics is trivially
free from relative short--distance singularities, i.e. singularities due
to their mutual interaction. In this case a Dirac--brane can (must) be used
to describe their dynamics, and the minimal--coupling Wess--Zumino term 
describing the mutual interaction is independent of the Dirac--brane 
{\it mod} $2\pi$, if Dirac's quantization condition holds.

If, on the other hand, the configuration is exceptional 
($\Sigma\neq \emptyset$) the two branes stay at zero distance. In this case 
the mutual interaction is plagued by short--distance singularities
and we will see that the canonical minimal $WZ$--term becomes 
{\it Dirac--brane--dependent}, i.e. it carries a (non--integer) 
Dirac--anomaly. In this letter we show that for such configurations the 
recently developed Chern--kernel approach \cite{LM,Chern,LMT} allows to 
write a new (manifestly) {\it Dirac--brane--independent} $WZ$--term.
However, this new $WZ$--term turns out to
be plagued by an inflow gravitational anomaly -- supported on  
$\Sigma$ -- if $d=0$ or $2$, while it is anomaly free if $d=1$ or $3$.
A crucial ingredient for its construction is the new ``descent--identity'' 
\eref{id}.

Eventually we show that for $d=0,2$ the inflow gravitational anomaly on 
$\Sigma$ is cancelled by the quantum anomaly produced by the fermions 
living on it. This means that the total classical + quantum effective action
is 1) free from gravitational anomalies and 2) Dirac--brane--independent.

The canonical and new $WZ$--terms differ by a local counterterm
supported on $\Sigma$, that maps the Dirac--anomaly in the gravitational
anomaly, whose construction will be outlined in the concluding section. 

For the anomaly cancellation mechanism of open $M2$--branes ending on 
$M5$--branes see \cite{LM,Mou}.

\section{Chern--kernels and Dirac--branes}

In presence of a closed $M2$-- and a closed $M5$--brane the Bianchi
identity and equation of motion for the four--form fieldstrength of 
eleven--dimensional Supergravity amount to \footnote{For simplicity we 
omit in \eref{2} the gravitational curvature 
polynomial $X_8$, which corrects eleven--dimensional Supergravity by the term 
$\int B X_8$ \cite{Tens.X8}.}
\ba
dH&=&J_5,\label{1}\\
d*H&=&{1\over2}HH+hJ_5+J_8,\label{2}
\ea
where $J_8$ $(J_5)$ is the $\delta$--function supported Poincar\`e--dual form
of the electric (magnetic) brane worldvolume $M2$ $(M5)$, i.e. its current;
$h=db+B|_{M5}$, where $b$ is the chiral two--form on $M5$ and $B$ is the
potential for $H$. The brane tensions are set to $T_{M2}=T_{M5}=2\pi$. 

In absence of the $M2$--brane ($J_8=0$) the basic ingredient for
the construction of a consistent $WZ$--term for this system (see \eref{wza}) 
is the Chern--kernel. 
We recall now briefly the essential features of this construction \cite{LMT}, 
concentrating on the main properties of the Chern--kernel, the details of 
the resulting $WZ$--term itself -- $S_{WZ}^A$ -- being unessential for what 
follows.  

To write an action for the system above one must first solve \eref{1} in terms
of a potential, introducing a four--form antiderivative $K$ for the 
magnetic current $J_5$, 
\be
dK=J_5, \quad H=dB+K.
\ee
This solution is subject to the transformations (called $Q$--transformations 
in the following)
\be
\label{Qtrans}
K^\prime=K+dQ,\quad B^\prime = B-Q, \quad Q|_{M5}=0. \label{4}
\ee
These transformations leave the curvatures $H$ and $h$ invariant, and  in 
writing an action one must ensure that this invariance -- $Q$--invariance --
remains preserved. 

As shown in \cite{LM,LMT}, the r.h.s. of \eref{2} becomes a well--defined 
closed form and there exists a $Q$--invariant action 
if one chooses as solution for $dK=J_5$ a four--form Chern--kernel,
 \be
\label{5}
K={1\over 4(4\pi)^2}\,\ve^{a_1\cdots a_5}\,\hat
y^{a_1}{\cal F}^{a_2a_3}{\cal F}^{a_4a_5},
\quad {\cal F}^{ab}=F^{ab}+D\yh^aD\yh^b,
 \ee
where the $y^a$ $(a=1,\cdots,5)$ are normal coordinates on $M5$, 
$\yh^a=y^a/|\vec y|$, $D\yh$ is the covariant differential w.r.t the normal
bundle $SO(5)$--connection $A$, and $F=dA+AA$ is its curvature. This kernel,
although being invariant near $M5$,
is not unique but subjected to the $Q$--transformations \eref{4} 
\cite{LM}. For what follows it is important to notice 
that $K$ is singular on the whole $M5$, 
because $\yh^a$ does not admit limit when $y^a$ goes to 0, while the 
three--form $Q$ is regular and has, actually, vanishing pullback  
on $M5$. The four--form $K$ can be seen as a kind of generalized 
Coulomb--like
field (inverse--power--like singularities), or also as an angular form 
\cite{FHMM}. We recall
also that in normal coordinates the $M5$--brane current reads
$J_5=dy^1\cdots dy^5\,\delta^5(y)$. 

In absence of the $M2$--brane one can then write down a $WZ$--term
giving rise to the eq. of motion \eref{2}. It is convenient to write
it as an integral over a twelve--dimensional manifold $M_{12}$ whose boundary  
is the eleven--dimensional target space $M_{11}$, of a {\it closed and
$Q$--invariant} twelve--form 
\footnote{Eventually, to get \eref{2} one has to 
take into account also the kinetic Born--Infeld--type action for $h$, see 
\cite{LM}.}, 
\be
\label{wza}
S_{WZ}^A=2\pi\int_{M_{12}}L_{12}^A,\quad
L_{12}^A={1\over 6}\,HHH+{1\over 2}\,h\,dh J_5
    +{1\over 24}\,P_7^{(0)} J_5,
\ee
where $P_8=dP_7^{(0)}$ is the second Pontrjagin form of the normal bundle
of $M5$. We remember that the property $dL_{12}^A=0$ ensures that,
in absence of topological obstructions, $S_{WZ}^A$ does not depend 
on the particular $M_{12}$ chosen. 
Writing it in this way the $WZ$ is manifestly $Q$--invariant, depending
only on $H$ and $h$, and in \cite{LMT} it has been shown that 
$L_{12}^A$ is a closed form and that $S_{WZ}^A$ cancels the residual
normal bundle $SO(5)$--anomaly localized on the $M5$--brane, \cite{Witten}.

As we observed already, for what follows the detailed form of 
$L_{12}^A$ is irrelevant; what is
crucial is that its consistency relies heavily on the presence of 
the Chern--kernel and on the corresponding solution $H=dB+K$ of the 
Bianchi--identity $dH=J_5$. 

In presence of an $M2$--brane, $J_8\neq 0$, one can write
the $WZ$--term as 
\be
S_{WZ}=S_{WZ}^A+S_{WZ}^B=
2\pi\int_{M_{12}}\left(L_{12}^A+L_{12}^B\right),\quad dL_{12}^B=0,
\quad L_{12}^B\quad Q-invariant,
\ee
where $L_{12}^B$ must describe 1) the minimal interaction of
the $M2$--brane with supergravity and 2) the mutual interaction between 
$M2$ and $M5$. The first interaction is canonical and corresponds to a 
contribution to $L_{12}^B$ given by $dB J_8 =d(BJ_8)$;
the presence of the second is needed because $dBJ_8$, although being
closed, is not $Q$--invariant. A $Q$--invariant completion could be achieved
by adding the  mutual interaction term $KJ_8$, leading to 
$dBJ_8+KJ_8=HJ_8$, but this is no longer a closed form. Eventually one should 
have 
\be
\label{dots}
L_{12}^B=HJ_8+\cdots,
\ee
where the missing terms have to be 1) $Q$--invariant, 2) $B$--independent and 
3) such that $L_{12}^B$ becomes a closed form. Our main problem consists 
therefore in figuring out to what the missing terms correspond to.

If $M2$ and $M5$ do not intersect there is, of course, a standard procedure
for writing down the missing terms above, that involves an (electric) 
Dirac--brane for $M2$ i.e. a four--surface $D_4$ whose boundary is $M2$, 
$\partial D_4=M2$. Denoting the 
$\delta$--function on $D_4$ with $W_7$, a seven--form, we have 
$$
J_8=dW_7,
$$
and one can perform the completion 
\be
L_{12}^B=HJ_8-J_5W_7=d\left(HW_7\right),
\ee
which satisfies all the above requirements. Of course, under a change of 
Dirac--brane the $WZ$--action 
\be
\label{Dirac}
S_{WZ}^B=2\pi \int_{M_{11}}HW_7 =2\pi \int_{D_4}H
\ee
changes by an integer multiple of $2\pi$, since $H$ has integer integrals
over any closed four--manifold that does not intersect $M5$ 
\footnote {For an alternative argument for Dirac--brane--independence, based 
on {\it integer} forms, see \cite{LM1}.}.

From a twelve--dimensional point of view independence
of the Dirac--brane is manifest since the Dirac--brane--dependent 
term $J_5W_7$ has integer integrals over arbitrary (closed or open)
manifolds. 

On the other hand, if the intersection manifold $\Sigma$ is non empty the 
$WZ$--action  $\int_{D_4}H$ becomes Dirac--brane--{\it dependent}. Indeed, 
under a change of Dirac--brane it changes by 
$$
\int_{D^\prime_4}H - \int_{D_4}H=\int_{S_4}H=\int_{S_4}K,
$$
where $S_4$ is a closed four--manifold. But since $M2$ intersects $M5$ also
$S_4$ intersects $M5$ and therefore part of the flux of $K$ stays in $S_4$
and part stays outside. This means that  $\int_{S_4}K$ is no longer integer,
and it represents a Dirac--anomaly.

From a twelve--dimensional point of view, the term $J_5W_7$
can no longer be used to make $L_{12}^B$ a closed form, because
if $\Sigma$ is non empty then the product $J_5W_7$ contains squares of 
$\delta$--functions ($\delta(x)\delta(x)$) and becomes ill--defined; 
this is a consequence of the non--vanishing
intersection of the normal bundles of the two branes, see below.

The canonical Dirac--brane construction must therefore be abandoned if 
$\Sigma\neq\emptyset$. In this case, since $L_{12}^B$ must be closed, 
the first step to find out what the 
missing terms in \eref{dots} may be, consists in computing 
$d(HJ_8)=d(KJ_8)$. Now, the product $KJ_8$ and its differential
in the sense of distributions are well defined even 
if $\Sigma\neq\emptyset$ 
(for $d\neq 3$) \footnote{The case $d=3$, corresponding
to $M2\subset M5$, is in some sense trivial and will be solved separately
below.}, but the point is that one is not allowed to 
apply Leibnitz's rule to compute it. In fact, the result obtained using 
naively this rule, i.e. $d(KJ_8)=J_5J_8$, contains squares
of $\delta$--functions -- for the same reasons as above --
and it is ill--defined. 

On the other hand, 
as we will show in the next section, the result 
of the evaluation of $d(KJ_8)$ in the sense of distributions is 
well--defined, and it has a simple interpretation if expressed in terms of 
normal bundles.

\section{Intersecting branes and normal bundles}

Suppose that $\Sigma=M2\cap M5$ is a closed manifold with dimension 
$d=0,1,2$ or 3, and introduce its current $J$
which is a closed $(11-d)$--form. For this case we will show
that the unknown terms in \eref{dots} can be deduced from a new kind
of ``descent--identity'' -- as such formulated in thirteen 
dimensions -- that involves the normal bundles of $M2$ and $M5$.

The normal bundles of $M2$, $M5$ and $\Sigma$, denoted by 
$N_{M2}$, $N_{M5}$ and $N_\Sigma$, 
carry respectively fibers of dimensions 8, 5 and $11-d$. 
On $\Sigma$ the bundles of $M2$ and $M5$ intersect to a bundle 
${\cal N}=N_{M2}\cap N_{M5}$, whose fiber is of dimension 
$n=5+8-(11-d)=d+2$, with structure group $SO(n)$. For example, if 
the intersection is just a point -- a $(-1)$--brane -- then $n=2$; and if 
$M2\subset M5$ then $n=5$ because in this case ${\cal N}=N_{M5}$. 

If $n$ is even we can define the Euler--form $\chi$ of the bundle
${\cal N}$, a form of degree $n$; if $n$ 
is odd we take $\chi$ to be zero by definition. The Euler--forms 
of interest are then
$$
\chi_2={1\over 4\pi}\,\ve^{r_1r_2}T^{r_1r_2},\quad 
\chi_4={1\over 2(4\pi)^2}\,\ve^{r_1r_2r_3r_4}T^{r_1r_2}T^{r_3r_4}, 
$$
where $T^{rs}$ is the curvature of ${\cal N}$. Our descent notations are
$\chi=d\chi^{(0)}$, $\delta \chi^{(0)}=d\chi^{(1)}$.

In going to thirteen dimensions we want to keep the degrees 
of the currents $J_8,J_5,J$ unchanged. This implies that the worldvolumes 
of $M2$, $M5$ and $\Sigma$ have to be extended respectively to
five--, eight,-- and $(d+2)$--dimensional manifolds. This keeps  
the dimensions of the normal bundles, in particular the dimension of 
${\cal N}$ and hence the degree of $\chi$, unchanged. In absence of 
topological obstructions
such extensions are always possible \cite{LMT}, and they were implicitly
understood in the twelve--dimensional construction of the previous 
section.

The result of the computation we referred to above amounts then to a 
descent--identity between thirteen--forms, involving the $\delta$--function
on $\Sigma$ and the Euler--form of ${\cal N}$,
\be
\label{id}
d(KJ_8)=J\,\chi, \quad  {\rm whenever}\quad M2\not\subset M5. 
\ee
The proof is given in the appendix. It is obvious
that $d(KJ_8)$ must be supported on $\Sigma$ and hence proportional 
to $J$; the proportionality factor $\chi$ follows then essentially for 
invariance reasons. It is understood, as said above, that for $d=1$ one takes 
$\chi=0$.
If $M2\subset M5$ $(d=3)$ the product $KJ_8$ is ill--defined; this case is   
in some sense trivial and it is solved separately below.

Our descent--identity is, actually, a {\it local} realization of the 
corresponding {\it cohomological} relation presented in \cite{CY}.

\section{Wess--Zumino action and anomaly cancellation}

Given the above identity it is easy to complete the twelve--form 
\eref{dots} to make it a {\it $Q$--invariant and closed} form ($d\neq 3$):
\be
\widetilde L_{12}^B=HJ_8-J\,\chi^{(0)},
\ee
where we introduced a standard Chern--Simons form through  
$\chi=d\chi^{(0)}$. The Wess--Zumino
actions for the four possible intersection manifolds of $M5$ and $M2$ can
then be written eventually as 
\be 
{1\over 2\pi}\widetilde S_{WZ}^B=\int_{M_{12}}\widetilde L_{12}^B=\int_{M2}B+
      \left\{\begin{array}{ll} 
      \int_{M_{12}}\left(KJ_8-J_{11}\,\chi_1^{(0)}\right), 
       &\quad  d=0,\\
      \int_{M_{12}}KJ_8,  & \quad d=1,\\
      \int_{M_{12}}\left(KJ_8-J_9\,\chi_3^{(0)}\right), 
       &\quad  d=2,\\
      0,&\quad d=3.
\end{array}\right.
\label{lista} 
\ee  
We recall that the terms $KJ_8$ are required for $Q$--invariance,
and that the terms with the Euler--forms are needed to ensure
independence of the particular $M_{12}$ chosen.  
For $d=0,2$ the Euler--form is non--vanishing, while for $d=1$ it 
vanishes. In this case $KJ_8$ is indeed a closed form, see \eref{id}.

For $d=3$ ($M2\subset M5$) the product $KJ_8=K|_{M2}J_8$ is ill--defined, 
because $K$ does not admit pullback on $M5$,
and \eref{id} is therefore not applicable. But in this case the term
$\int_{M2}B$ is, actually, $Q$--invariant. Indeed, under a $Q$--transformation
one would have 
$\int_{M2}\left(B^\prime -B\right)=-\int_{M2}Q$, and this vanishes because
$Q$ vanishes on the whole $M5$, see \eref{Qtrans}. In other words,
for $M2\subset M5$ the form $\widetilde L_{12}^B=dBJ_8$ is already 
$Q$--invariant and closed; this explains the fourth line in \eref{lista}.

From the list \eref{lista} one sees that for $d=0,2$ the $WZ$--action
is plagued by a gravitational anomaly supported on the intersection manifold 
$\Sigma$, $\delta \widetilde S_{WZ}^B=-2\pi\int_{\Sigma}\chi^{(1)}$,
corresponding to the inflow anomaly--polynomial
\be
\label{inflow}
-2\pi\,\chi,
\ee
while for $d=1,3$ it is anomaly free.

On the other hand on $\Sigma$ there are also fermions 
living, coming from 
the common reduction of the 32--component spinors $\vartheta^\alpha$, 
living on $M5$ and on $M2$. If $d$ is even, these fermions are a section 
of the chiral spinor bundle
lifted from $T(\Sigma)\oplus {\cal N}$, where $T(\Sigma)$ is the
tangent bundle to 
$\Sigma$, and ${\cal N}$ is the intersection of the normal bundels, as above.
For such fermions the anomaly can be computed as in
\cite{CY}, and the resulting polynomial reads
\be
\label{index}
2\pi\left(ch[S^+_{\cal N}]-ch[S^-_{\cal N}]\right)\hat A\,[T(\Sigma)]
=2\pi\,{\hat A\,[T(\Sigma)]\over \hat A\,[{\cal N}]}\, \chi,
\ee
where $ch$ indicates the Chern character, $S^{\pm}_{\cal N}$ is the spin
bundle lifted from ${\cal N}$ with $\pm$ chirality, and $\hat A$ is the roof
genus. From this polynomial one has to extract the two--form part for
$d=0$, and the four--form part for $d=2$. Since the Euler form 
is already a form of degree two and four respectively, the roof genera above
contribute both with unity and the anomaly polynomial 
reduces precisely to $2\pi \chi$, cancelling the inflow.

This represents a quantum consistency check of the intersecting
$M2$/$M5$ configurations considered in this paper.

\section{Concluding remarks}

The anomaly cancellation mechanism presented in this letter has a 
transparent meaning for $d=2$, where $\Sigma$ is the worldvolume
of a closed string. 

For $d=0$ the intersection manifold $\Sigma$ is just a point $P$ and 
represents an instanton. In this case the bundle 
${\cal N}$ is a two--plane centered on $P$, with structure group
$SO(2)$ and abelian connection $W^{rs}$, 
and the inflow anomaly $-2\pi\int_\Sigma \chi^{(1)}$   
reduces to $-\Lambda(P)$, where $\delta W^{rs}=d(\ve^{rs}\Lambda)$.
This anomaly has a clear meaning since the 
normal bundle transformations of ${\cal N}$ correspond just to rotations
around $P$ in the two--plane centered at $P$, 
and $\Lambda(P)$ is the variation of the polar angle $\varphi$ of that 
plane, $\delta\varphi= \Lambda(P)$. What is more obscure is the meaning 
of (chiral) fermionic degrees of freedom on an instanton and the appearance
of the corresponding quantum anomaly. In lack of this insight, above
we took simply advantage from the fact that the index formula 
\eref{index} makes sense also for $d=0$. 

Our strategy for constructing a consistent interaction between intersecting
$M2$-- and $M5$--branes assumes that the branes intersect ``strictly'', i.e.
that they stay strictly at zero distance. An alternative strategy for 
describing intersecting branes would be to introduce a framing regularization,
where the branes are moved at a finite distance $\ve$ from each other. 
For each finite $\ve$ the branes are non--intersecting and one could
introduce consistently a Dirac--brane to describe their interaction, as 
explained in section two, see \eref{Dirac}. However, for $\ve \rightarrow 0$ 
this $WZ$--term, although remaining finite, would become
Dirac--brane--dependent, as explained in the text.

Keeping then the branes strictly intersecting, there are two ways
for writing a classical action. The first is $\widetilde S_{WZ}^B$ given 
in \eref{lista}: it is (manifestly) Dirac--brane--independent but 
carries a gravitational
anomaly (for $d=0,2$). The second is $S_{WZ}^B$ given in \eref{Dirac}: it is 
(manifestly) free from gravitational anomalies but it is plagued by a 
Dirac--anomaly. This means that there exists a local (in the 
sense of ``Wess--Zumino'') counterterm that maps the Dirac--anomaly
in a gravitational anomaly and vice versa. Its implicit construction
goes along the following lines. Starting point is an identity
similar to \eref{id},
\be
\label{new}
d\left(KW_7\right)=KJ_8-J\Phi,
\ee
for some $(d+1)$--form $\Phi$, which can be proven using -- for example -- the
regularizations given in the appendix
of \cite{LM}; again, one is not allowed to use naively 
Leibnitz's rule. For $d=3$ the term $KJ_8$ in this identity has to be 
replaced by $0$. The form $\Phi$, supported on $\Sigma$, is diffeomorphism 
invariant, but depends on $W_7$ i.e. on the Dirac--brane $D_4$. 
Using \eref{id} together with \eref{new} one gets
$$  
d\,\Phi=\chi \Rightarrow \Phi-\chi^{(0)}=d\omega,
$$
for some $d$--form $\omega$ on $\Sigma$; for $d$ odd $\Phi$ is thus a closed 
form. It is then immediately seen that 
$$
\widetilde S_{WZ}^B= S_{WZ}^B +2\pi\int_\Sigma \omega.
$$
The counterterm we searched for is $\int_\Sigma\omega$ and it is 
supported on $\Sigma$, as one may have expected. 
This proves in particular that the Dirac--anomaly itself is 
supported on $\Sigma$, in agreement with the fact that if 
$\Sigma =\emptyset$ then there is no Dirac--anomaly at all.

The configurations we have considered in this letter are exceptional
in that, a priori, a small perturbation makes the two branes again 
non--intersecting. The stability of these configurations can, 
however, be inferred from the existence of their classical--solution
(semi--localized) counterparts of $D=11$ Sugra, mentioned in the introduction,
whose stability is guaranteed by supersymmetry. There exist indeed 
solutions for $d=3$, preserving 1/2 susy \cite{Izq}, and solutions for 
$d=2$, preserving 1/4 susy \cite{Tsey1,Gaun}. To our knowledge no
solutions for $d=1$ or $d=0$ are yet known. The results of the present
paper, indicating that intersecting $M2/M5$ configurations
are quantum consistent for any value of $d$, suggest that also for $d=0,1$ 
supersymmetric classical solutions may exist. A dimensional reduction of
the complete interacting system Sugra+$M2$+$M5$ to ten dimensions, 
analogous to the one of \cite{CL}, may help to answer this question.

\bigskip

\paragraph{Acknowledgements.}
The author thanks M. Cariglia and P.A. Marchetti for useful discussions.
This work is supported in part by the European Community's Human
Potential Programme under contract HPRN-CT-2000-00131 Quantum
Spacetime.

\section{Appendix: proof of the descent--identity}

Since $J_8$ is the $\delta$--function on $M2$, the first step in 
proving \eref{id} consists  in evaluating $K$ 
restricted to $M2$.  After that one can compute
\be
\label{eval}
d(KJ_8)=d\left(K|_{M2}J_8\right)=d\left(K|_{M2}\right)J_8.
\ee
Since away from $M5$ $K$ is a closed form, the pullback $K|_{M2}$
is closed a part from (possible) $\delta$--function contributions 
supported on $\Sigma$. This means that it is sufficient to evaluate 
$K|_{M2}$ in the vicinity of $\Sigma$. 
On $\Sigma$ the normal bundle of $M5$ has 
$n$ coordinates in common with the normal bundle of $M2$ -- precisely
the ones of ${\cal N}$ -- so we can
split the normal coordinates of $M5$ as $y^a=(y^r,y^i)$, $(r=1,\cdots,n)$, 
$(i=n+1,\cdots,5)$.
Near $\Sigma$ we have $y^r=0$, while the $y^i$ become $3-d=5-n$ normal 
coordinates for $\Sigma$ with respect to $M2$. Correspondingly 
the $SO(n)$--connection $W$ of ${\cal N}$ is embedded 
in the $SO(5)$--connection $A$ according to $A^{rs}=W^{rs}$, $T=dW+WW$.

We perform now the evaluation of $K|_{M2}$ near $\Sigma$, i.e. for $y^r=0$
and keeping only terms that can give $\delta$--function contributions when
applying the differential, for each case separately.

${\bf d=0.}$ 
In this case $\Sigma$ is a point in $D=11$, corresponding to
a two--dimensional surface in $D=13$, and  its current is an eleven--form 
$J=J_{11}$. The 
fiber of ${\cal N}$ is two--dimensional ($n=2$), with structure--group 
$SO(2)$ and Euler--form $\chi_2(T)$. The identity to be proved is therefore
\be
\label{again}
d(KJ_8)=J_{11}\,\chi_2(T).
\ee
$\delta$--function contributions localized at $\Sigma$ are supported in
$y^i=0$, $(i=3,4,5)$ and they can arise from the angular form 
$$
K_0={1\over 8\pi}\,\ve^{ijk}\yh^id\yh^jd\yh^k,
$$ 
since $dK_0=d^3y\,\delta^3(y)$. It is then easy to evaluate \eref{5} for 
$y^r=0$, $(r=1,2)$ and to extract the contribution proportional to 
$K_0$,
$$
K|_{M2}={1\over 4\pi}\,\ve^{rs}
\left(T^{rs}+\left(\delta^{ij}-3\,\yh^i\yh^j\right)A^{ir}A^{js}\right) K_0.
$$
Since one has $d[(\delta^{ij}-3\yh^i\yh^j)K_0]=0$, when taking the 
differential only the first term contributes with a $\delta$--function 
and one gets 
$$
d\left(K|_{M2}\right)=d^3y\,\delta^3(y)\,\chi_2(T).
$$
\eref{again} follows then from \eref{eval} and from 
$d^3y\,\delta^3(y)\,J_8=J_{11}$.

${\bf d=1.}$ 
In this case $\Sigma$ is a worldline in $D=11$ and its current is a 
ten--form $J=J_{10}$. The 
fiber of ${\cal N}$ is three--dimensional ($n=3$) and its
Euler--form vanishes. The descent--identity reduces therefore to
$$
d(KJ_8)=0.
$$
As above one should extract from $K|_{M2}$, taken at $y^r=0$,
$(r=1,2,3)$, contributions proportional to the 
angular form, that is now 
$K_0={1\over 2\pi}\ve^{ij}\yh^id\yh^j$, $(i,j=4,5)$,
$dK_0=d^2y\delta^2(y)$. But since $K$ contains only odd powers of 
$\yh$, and $K_0$ is even in $\yh$, in $K|_{M2}$ the angular form 
$K_0$ appears always multiplied by odd powers of $\yh$, and taking the 
differential the current $d^2y\,\delta^2(y)$ can never show up. This implies 
that $d\left(K|_{M2}\right)=0$.
 
${\bf d=2.}$ 
In this case $\Sigma$ is a two--surface in $D=11$, and  its current is a
nine--form $J=J_9$. The 
fiber of ${\cal N}$ is four--dimensional ($n=4$) with structure--group 
$SO(4)$ and Euler--form $\chi_4(T)$. The identity becomes then 
$$
d(KJ_8)=J_9\,\chi_4(T).
$$
In this case one has $r=1,2,3,4$ and $i=5$, and it is straightforward
to evaluate \eref{5} at $y^r=0$, 
$$ 
K|_{M2}={1\over 4(4\pi)^2}\,{y^5\over |y^5|}\,\ve^{r_1r_2r_3r_4}\,
T^{r_1r_2}T^{r_3r_4}={y^5\over 2|y^5|}\,\chi_4(T).
$$
The differential of the ``angular--form'' is here simply $d(y^5/2|y^5|)=
dy^5\delta(y^5)$, and one gets
$$
d\left(K|_{M2}\right)=dy^5\delta(y^5)\,\chi_4(T).
$$
One concludes then as in the case $d=0$. 

${\bf d=3.}$ In this case we have $n=5$, $J=J_8$ and $\chi=0$,
and the r.h.s. of \eref{id} vanishes. On the other hand, since 
$\Sigma=M2\subset M5$, the pullback  $K|_{M2}$
would require to evaluate $K$ for $y^a\rightarrow 0$. But this limit depends
on the direction $\yh^a=V^a(\sigma)$ one chooses to approach $M2$ at each 
point $\sigma$, and $K|_{M2}$ is ill--defined. Notice however, that if one
performs the limit along an arbitrary but fixed vector field $V^a(\sigma)$, 
then the resulting four--form $(K|_{M2})^V$ can be shown to be closed. 
But such a definition saves the identity \eref{id} {\it only formally}, 
because $K|_{M2}$, and hence $KJ_8$, acquires a dependence on an unphysical 
vector field, and it can not be used in the $WZ$--action.



\end{document}